\documentclass[10pt,pra,twocolumn,showpacs,floatfix]{revtex4}
\usepackage{amsfonts}
\usepackage{amssymb}
\usepackage{amsmath}
\usepackage[dvips]{graphicx}

\begin{document}

\title{Quantum entanglement swapping between two multipartite entangled
states}
\author{Xiaolong Su$^{1,2}$}
\email{suxl@sxu.edu.cn}
\author{Caixing Tian$^{1,2}$}
\author{Xiaowei Deng$^{1,2}$}
\author{Qiang Li$^{1,2}$}
\author{Changde Xie$^{1,2}$}
\author{Kunchi Peng$^{1,2}$}
\email{kcpeng@sxu.edu.cn}
\affiliation{$^{1}$State Key Laboratory of Quantum Optics and Quantum Optics Devices, \\
Institute of Opto-Electronics, Shanxi University, Taiyuan, 030006, People's
Republic of China\\
$^{2}$Collaborative Innovation Center of Extreme Optics, Shanxi University,\\
Taiyuan,Shanxi 030006, People's Republic of China}

\begin{abstract}
Quantum entanglement swapping is one of the most promising ways to realize
the quantum connection among local quantum nodes. In this Letter, we present
an experimental demonstration of the entanglement swapping between two
independent multipartite entangled states, each of which involves a
tripartite Greenberger-Horne-Zeilinger (GHZ) entangled state of an optical
field. The entanglement swapping is implemented deterministically by means
of a joint measurement on two optical modes coming from the two multipartite
entangled states respectively and the classical feedforward of the
measurement results. After entanglement swapping the two independent
multipartite entangled states are merged into a large entangled state in
which all unmeasured quantum modes are entangled. The entanglement swapping
between a tripartite GHZ state and an Einstein-Podolsky-Rosen entangled
state is also demonstrated and the dependence of the resultant entanglement
on transmission loss is investigated. The presented experiment provides a
feasible technical reference for constructing more complicated quantum
networks.
\end{abstract}

\maketitle

Multipartite entangled states play essential roles in quantum computation
and quantum networks. Cluster states, a type of multipartite entangled
states, are basic quantum resources for one-way quantum computation \cite%
{Briegel,Zhang2006}. Based on a prepared large scale cluster state, one-way
quantum computation can be implemented by measurement and feedforward of the
measured results \cite{Raussendorf2001,Menicucci2006,Ukai2011,Su2013}. It
has been demonstrated that a local quantum network can be built by
distributing a multipartite entangled state among quantum nodes \cite%
{Loock,Hide,Jing,Jona}. If we have two space-separated local quantum
networks built by two independent multipartite entangled states,
respectively, how can we establish entanglement between the quantum nodes in
the two local quantum networks? It has been proposed that a large scale
cluster state can be generated by the fusion of small scale cluster states,
which is completed by means of linear optical elements \cite{Browne2005}.
The shaping of a larger cluster state to a smaller one according to the
requirement for one-way quantum computation has been demonstrated \cite{Miwa}%
. Another feasible method of merging two multipartite entangled states into
one larger multipartite entangled state is quantum entanglement swapping 
\cite{Bose}, which has been proposed to build a global quantum network of
clocks that may allow the construction of a real-time single international
time scale (world clock) with unprecedented stability and accuracy \cite%
{Komar}.

Quantum teleportation enables transportation of an unknown quantum state to
a remote station by using an entangled state as the quantum resource. Up to
now, long distance quantum teleportation of single photons over 100 km has
been experimentally demonstrated \cite{Pan1,Ma,Nam}. Quantum entanglement
swapping, which makes two independent quantum entangled states become
entangled without direct interaction, is an important technique in building
quantum communication networks \cite%
{Zukowski,Pan,Sciarrino,Riedmatten,Ralph,Tan,Loock2,Jia,Takei}. Quantum
entanglement swapping is also known as quantum teleportation of entangled
states \cite{Loock2}. It was originally proposed and demonstrated in
discrete-variable optical systems \cite{Zukowski,Pan}, and then it was
extended to continuous-variable systems \cite{Ralph,Tan,Loock2,Jia,Takei}.
Recently, entanglement swapping between discrete and continuous variable
systems has been demonstrated \cite{Takeda}, which shows the power of hybrid
quantum information processing \cite{Andersen}. The entanglement swapping
among three two-photon Einstein-Podolsky-Rosen (EPR) entangled states has
been used to generate a Greenberger-Horne-Zeilinger (GHZ) state \cite{Lu2009}%
. However, quantum entanglement swapping between two multipartite entangled
states, each of which involves more than two subsystems, has not been
demonstrated.

\begin{figure*}[tbp]
\begin{center}
\includegraphics[width=145mm]{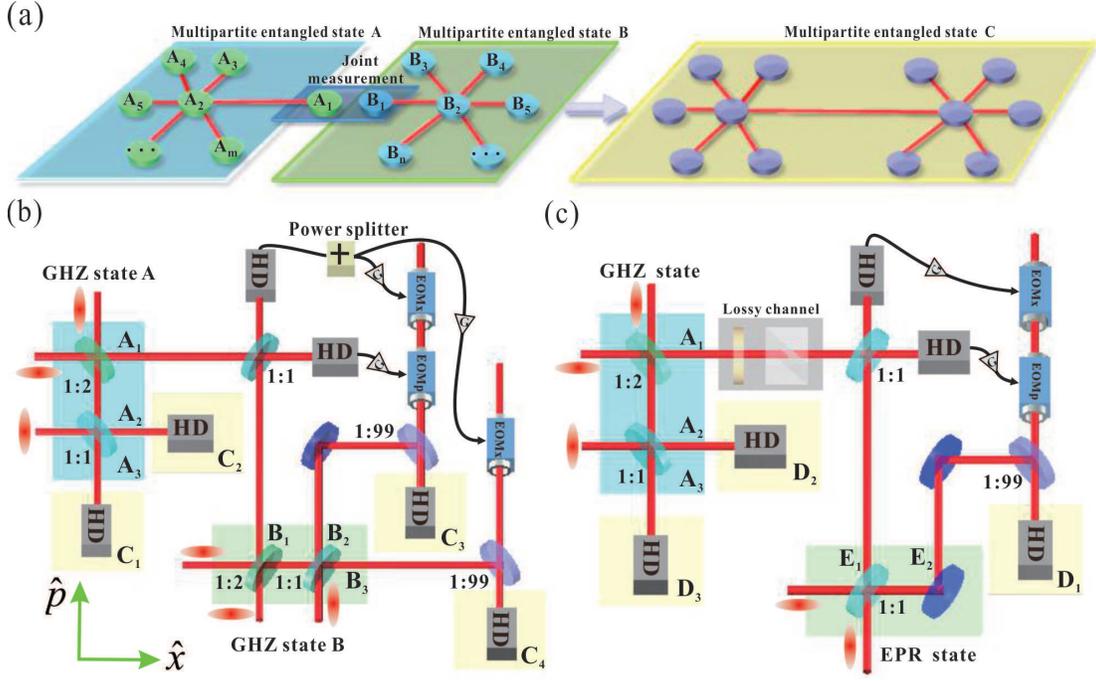}
\end{center}
\caption{Schematic of principle and experimental setup. (a) Schematic of
entanglement swapping between two multipartite entangled states \emph{A} and 
\emph{B}. The joint measurement is performed on optical modes $A_{1}$ and $%
B_{1}$ coming from two multipartite entangled states \emph{A} and \emph{B},
respectively. The measurement results are fedforward to the remained quantum
modes of multipartite entangled state \emph{B}. Then a new multipartite
entangled state \emph{C} is obtained after entanglement swapping. (b) The
schematic of the experimental setup for the entanglement swapping between
two tripartite GHZ entangled states. (c) The schematic of the experimental
setup for the entanglement swapping between a tripartite GHZ state and an
EPR entangled state. The lossy channel is simulated by a half wave plate and
a polarization beam splitter. EOM, electro-optic modulator; HD, homodyne
detector. The power splitter is used to split the output photocurrent from
the homodyne detector.}
\end{figure*}

In this Letter, we present the first experimental demonstration of
deterministic entanglement swapping between two multipartite entangled
states of light. Two multipartite entangled states \emph{A} and \emph{B},
consisting of $m$ ($m\geqslant 2$) and $n$ ($n\geqslant 2$) optical modes,
respectively, are merged into a larger multipartite entangled state \emph{C}
as shown in Fig. 1(a). In order to establish entanglement between the two
multipartite entangled states, optical mode\emph{\ A}$_{1}$ is sent to the
multipartite entangled state\emph{\ B} through a quantum channel. A joint
measurement on modes\textit{\ }\emph{A}$_{1}$\emph{\ }and \emph{B}$_{1}$%
\emph{\ }is implemented and then the measured results are fed forward to the
remaining optical modes of state\emph{\ B}. The feedforward schemes of
measurement results in classical channels depend on the types of quantum
correlation of the multipartite entangled state, which are more complex than
that for the traditional entanglement swapping between two EPR entangled
states. By quantum entanglement swapping, two multipartite entangled states
consisting of $m$ and $n$ quantum nodes, respectively, can be merged into a
new larger multipartite entangled state consisting of $m+n-2$ quantum modes,
since two modes (\emph{A}$_{1}$\emph{\ }and\emph{\ B}$_{1}$) have been
measured [Fig. 1(a)].

As an example, we experimentally demonstrate the entanglement swapping
between two independent tripartite GHZ entangled states consisting of three
quantum modes $A_{1}$, $A_{2}$, $A_{3}$ and $B_{1}$, $B_{2}$, $B_{3}$,
respectively [Fig. 1(b)]. Each of the tripartite entangled state of optical
field is obtained by combining three squeezed states of light with $-5.90$
dB squeezing and $9.84$ dB anti-squeezing on two optical beam splitters with
transmissivity of $1/3$ and $1/2$, respectively [see APPENDIX A]. The
amplitude and phase quadratures of an optical mode $\hat{a}$ are expressed
by $\hat{x}_{a}=\hat{a}+\hat{a}^{\dagger }$ and $\hat{p}_{a}=(\hat{a}-\hat{a}%
^{\dagger })/i$, respectively. The correlation variances between the
amplitude and phase quadratures of three modes\textit{\ }\emph{A}$_{1}$\emph{%
\ (B}$_{1}$\emph{), A}$_{2}$\emph{\ (B}$_{2}$\emph{) and A}$_{3}$\emph{\ (B}$%
_{3}$\emph{)}\textit{\ }of a tripartite entangled state are expressed by $%
\Delta ^{2}\left( \hat{x}_{A_{1}(B_{1})}-\hat{x}_{A_{2}(B_{2})}\right)
=\Delta ^{2}\left( \hat{x}_{A_{2}(B_{2})}-\hat{x}_{A_{3}(B_{3})}\right)
=\Delta ^{2}\left( \hat{x}_{A_{1}(B_{1})}-\hat{x}_{A_{3}(B_{3})}\right)
=2e^{-2r}$ and $\Delta ^{2}\left( \hat{p}_{A_{1}(B_{1})}+\hat{p}%
_{A_{2}(B_{2})}+\hat{p}_{A_{3}(B_{3})}\right) =3e^{-2r}$, respectively,
where the subscripts correspond to different optical modes and $r$\ is the
squeezing parameter ($r=0$\ and $r=+\infty $\ correspond to no squeezing and
the ideally perfect squeezing, respectively). We have suggested that all
three squeezed states have identical squeezing and the requirement is easy
to be fulfilled in experiments \cite{Jia}.

The $\hat{x}$-squeezed and $\hat{p}$-squeezed states are produced by
nondegenerate optical parametric amplifiers (NOPAs) pumped by a common laser
source, which is a continuous wave intracavity frequency-doubled and
frequency-stabilized Nd:YAP/LBO (Nd-doped YAlO$_{3}$ perorskite - lithium
triborate) laser. The output fundamental wave at 1080 nm wavelength is used
for the injected signals of NOPAs and the local oscillators of homodyne
detectors. The second-harmonic wave at 540 nm wavelength serves as the pump
field of the NOPAs, in which through an intracavity
frequency-down-conversion process a pair of signal and idler modes with the
identical frequency at 1080 nm and the orthogonal polarizations are
generated. Each of the NOPAs consists of an $\alpha $-cut type-II potassium
titanyl phosphate (KTP) crystal and a concave mirror. The front face of the
KTP crystal is coated for the input coupler and the concave mirror serves as
the output coupler, which is mounted on a piezoelectric transducer to
actively lock the cavity length of NOPAs on resonance with the injected
signal at $1080$ nm. The transmissivities of the front face of the KTP
crystal at 540 nm and 1080 nm are $21.2\%$ and $0.04\%$, respectively. The
end face of the KTP crystal is cut to 1$%
%TCIMACRO{\U{b0}}%
%BeginExpansion
{{}^\circ}%
%EndExpansion
$ along the\textit{\ y-z} plane of the crystal and is antireflection coated
for both 1080 nm and 540 nm \cite{Zhou}. The transmissivities of the output
coupler at 540 nm and 1080 nm are $0.5\%$ and $12.5\%$, respectively. In our
experiment, NOPAs are operated at the parametric deamplification situation 
\cite{Zhou,Su2007}. Under this condition, the coupled modes at +45$^{\circ }$
and -45$^{\circ }$ polarization directions are the $\hat{x}$-squeezed and $%
\hat{p}$-squeezed states, respectively \cite{Su2007}.

The joint measurement is performed on modes $A_{1}$ and $B_{1}$ by a beam
splitter and two homodyne detectors. The measurement results are fed forward
to the remaining modes in multipartite entangled state \emph{B} through
classical channels, in which $G(\hat{x}_{A_{1}}-\hat{x}_{B_{1}})$ and $G(%
\hat{p}_{A_{1}}+\hat{p}_{B_{1}})$ are fed forward to the amplitude and phase
quadratures of mode $B_{2}$, and $G(\hat{x}_{A_{1}}-\hat{x}_{B_{1}})$ is fed
forward to the amplitude quadrature of mode $B_{3}$, respectively, where $G$
is the gain in the classical channels [see APPENDIX B]. The displacement
operations are performed by using electro-optical modulators (EOMs) and
highly reflecting mirrors (1:99 beam splitters). After entanglement
swapping, a new larger multipartite entangled state consisting of four
quantum modes ($C_{1}$, $C_{2}$, $C_{3}$, and $C_{4}$) is obtained.

The quantum entanglement swapping between a tripartite GHZ state ($m=3$) and
an EPR state ($n=2$) is also realized, where a new multipartite entangled
state \emph{D} involving three quantum modes ($D_{1}$, $D_{2}$ and $D_{3}$)
is obtained [Fig. 1(c)]. The EPR entangled state is prepared by coupling two
squeezed states of light on an optical beam splitter with transmissivity of $%
1/2$. The joint measurement results on modes $A_{1}$ and $E_{1}$, $G(\hat{x}%
_{A_{1}}-\hat{x}_{E_{1}})$ and $G(\hat{p}_{A_{1}}+\hat{p}_{E_{1}})$, are fed
forward to the amplitude and phase quadratures of mode $E_{2}$ through
classical channels [see APPENDIX C].

The gain in classical channels is an essential experimental parameter in
entanglement swapping \cite{Tan,Loock2,Jia}. The optimal gain $G=0.95$ for
classical channels is applied in the experiment, which reduces the demand
for the initial squeezing at the maximal extent. Here, the unit gain is not
selected because if $G=1$ the required squeezing level for obtaining the
resultant entangled state is higher than that of $G=0.95$ [see APPENDIX B
and C].

\begin{figure}[tbp]
\begin{center}
\includegraphics[width=85mm]{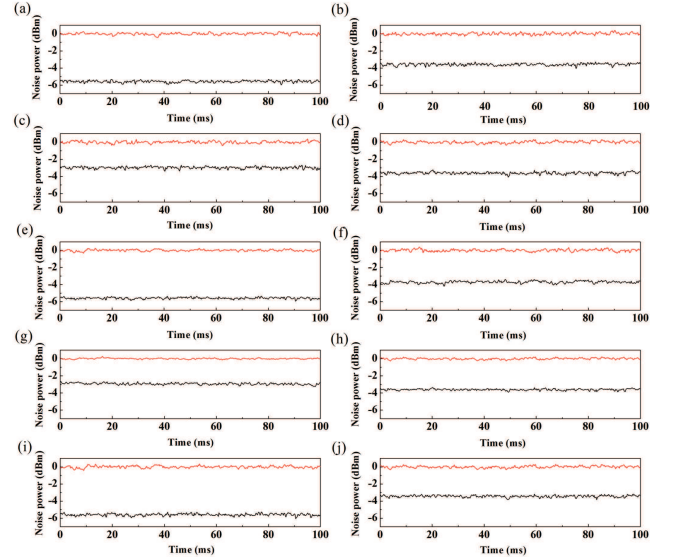}
\end{center}
\caption{The measured correlation noises of the output modes. (a)-(f) are
the correlation noises for the multipartite entangled state \emph{C}, which
are $\Delta ^{2}(\hat{x}_{C_{1}}-\hat{x}_{C_{2}})=$ $-5.57\pm 0.12$\ dB, $%
\Delta ^{2}(\hat{p}_{C_{1}}+\hat{p}_{C_{2}}+g_{1}\hat{p}_{C_{3}}+g_{2}\hat{p}%
_{C_{4}})=-3.58\pm 0.13$ dB, $\Delta ^{2}(\hat{x}_{C_{2}}-\hat{x}%
_{C_{3}})=-2.97\pm 0.13$ dB, $\Delta ^{2}(g_{3}\hat{p}_{C_{1}}+\hat{p}%
_{C_{2}}+\hat{p}_{C_{3}}+g_{4}\hat{p}_{C_{4}})=-3.61\pm 0.13$ dB, $\Delta
^{2}(\hat{x}_{C_{3}}-\hat{x}_{C_{4}})=-5.59\pm 0.10$ dB, $\Delta ^{2}(g_{5}%
\hat{p}_{C_{1}}+g_{6}\hat{p}_{C_{2}}+\hat{p}_{C_{3}}+\hat{p}%
_{C_{4}})=-3.71\pm 0.13$ dB, respectively. (g)-(j) are the correlation
noises for the multipartite entangled state \emph{D}, which are $\Delta ^{2}(%
\hat{x}_{D_{1}}-\hat{x}_{D_{2}})=-2.93\pm 0.11$ dB, $\Delta ^{2}(\hat{p}%
_{D_{1}}+\hat{p}_{D_{2}}+g_{7}\hat{p}_{D_{3}})=$ $-3.61\pm 0.09$ dB, $\Delta
^{2}(\hat{x}_{D_{2}}-\hat{x}_{D_{3}})=-5.59\pm 0.13$ dB, $\Delta ^{2}(g_{8}%
\hat{p}_{D_{1}}+\hat{p}_{D_{2}}+\hat{p}_{D_{3}})=-3.43\pm 0.12$ dB,
respectively. The red and black lines correspond to the shot noise level and
correlation noises, respectively. Measurement frequency is 3 MHz, parameters
of the spectrum analyzer: resolution bandwidth is 30 kHz, and video
bandwidth is 300 Hz.}
\end{figure}

\begin{figure*}[tbp]
\begin{center}
\includegraphics[width=150mm]{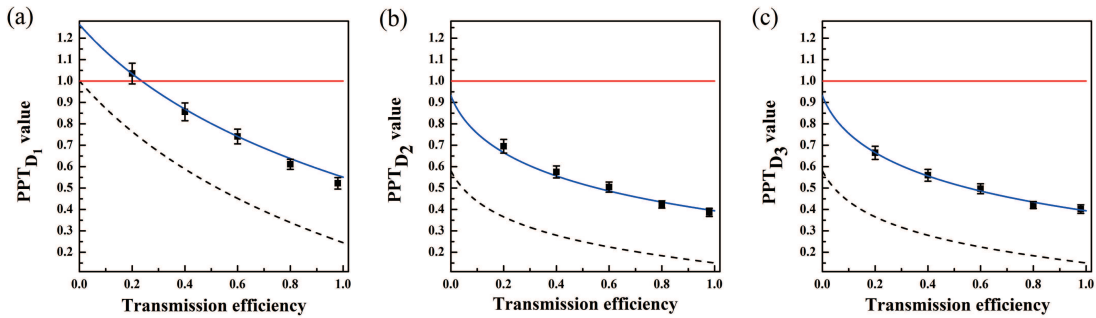}
\end{center}
\caption{The distributed entanglement in a lossy channel. (a)-(c), The PPT
values PPT$_{\text{\emph{D}}_{1}}$, PPT$_{\text{\emph{D}}_{2}}$ and PPT$_{%
\text{\emph{D}}_{3}}$ represent the different splittings for the ($%
D_{1}|D_{2}D_{3}$), ($D_{2}|D_{1}D_{3}$), and ($D_{3}|D_{1}D_{2}$),
respectively, which measure the inseparability of one mode with the other
two modes. The PPT values in a lossy channel (blue lines) are all below the
boundary (red lines) when channel efficiency is higher than 0.24. The dashed
lines are the obtained PPT values with $-10.9$ dB squeezing, which shows
that the obtained tripartite entanglement can be robust against loss in the
quantum channel. The black dots represent the experimental data. Error bars
represent $\pm $1 standard deviation and are obtained based on the
statistics of the measured noise variances.}
\end{figure*}

Figure 2 shows the measured quantum correlation variances of the output
states. The quantum entanglement among the output modes of multipartite
entangled state \emph{C} is verified by the inseparability criteria for a
four-mode GHZ entangled state, which are \cite{LF} 
\begin{eqnarray}
\Delta ^{2}(\hat{x}_{C_{1}}-\hat{x}_{C_{2}})+\Delta ^{2}(\hat{p}_{C_{1}}+%
\hat{p}_{C_{2}}+g_{1}\hat{p}_{C_{3}}+g_{2}\hat{p}_{C_{4}}) &<&4  \notag \\
\Delta ^{2}(\hat{x}_{C_{2}}-\hat{x}_{C_{3}})+\Delta ^{2}(g_{3}\hat{p}%
_{C_{1}}+\hat{p}_{C_{2}}+\hat{p}_{C_{3}}+g_{4}\hat{p}_{C_{4}}) &<&4  \notag
\\
\Delta ^{2}(\hat{x}_{C_{3}}-\hat{x}_{C_{4}})+\Delta ^{2}(g_{5}\hat{p}%
_{C_{1}}+g_{6}\hat{p}_{C_{2}}+\hat{p}_{C_{3}}+\hat{p}_{C_{4}}) &<&4  \notag
\\
&&  \label{E4}
\end{eqnarray}%
where $g_{i}$ $(i=1,2,...,6)$ is the optimal gain used to minimize the
correlation variances at the left-hand sides of Eq. (\ref{E4}). The value 4
at the right-hand sides of Eq. (\ref{E4}) is the corresponding boundary for
inseparability.\textbf{\ }When all correlation variances at the left-hand
sides of Eq. (\ref{E4}) are smaller than 4, four modes $C_{1}$, $C_{2}$, $%
C_{3}$, and $C_{4}$\ are entangled.\ From the measured results shown in
Figs. 2(a)-2(f), we can calculate the combinations of correlation variances
at the left-hand sides of the three inequalities, which are $2.10\pm 0.06$, $%
2.65\pm 0.08$, and $2.06\pm 0.06$ with $g_{1}=0.90$, $g_{2}=0.84$, $g_{3}=$ $%
g_{4}=0.94$, and $g_{5}=g_{6}=0.88$, respectively. The satisfaction of the
inseparability criteria of the four-mode GHZ state confirms the success of
quantum entanglement swapping between two tripartite GHZ entangled states of
light.

The inseparability criteria for a three-mode GHZ entangled state established
in the entanglement swapping between the tripartite entangled state and the
EPR entangled state are given by 
\begin{eqnarray}
\Delta ^{2}(\hat{x}_{D_{1}}-\hat{x}_{D_{2}})+\Delta ^{2}(\hat{p}_{D_{1}}+%
\hat{p}_{D_{2}}+g_{7}\hat{p}_{D_{3}}) &<&4  \label{E3} \\
\Delta ^{2}(\hat{x}_{D_{2}}-\hat{x}_{D_{3}})+\Delta ^{2}(g_{8}\hat{p}%
_{D_{1}}+\hat{p}_{D_{2}}+\hat{p}_{D_{3}}) &<&4  \notag
\end{eqnarray}%
where $g_{j}$ $(j=7,8)$ is the optimal gain used to minimize the correlation
variances at the left-hand sides of Eq. (\ref{E3}). From the measured
results shown in Figs. 2(g)-2(j), we obtain the values at the left-hand
sides of Eq. (2), which are $2.27\pm 0.06$ and $1.85\pm 0.05$ with $%
g_{7}=0.94$, $g_{8}=0.93$, respectively. The values are smaller than $4$ and
thus demonstrate the success of entanglement swapping between the tripartite
entangled state and the EPR entangled state.

We also consider the feasibility of completing entanglement swapping in a
real quantum communication network. In quantum communication, the losses and
noises in quantum channels lead to decoherence of quantum states and the
distributed entanglement will degrade (even disappear) by the unavoidable
decoherence. We simulate the loss in real quantum channels by using a half
wave plate and a polarization beam splitter as shown in Fig. 1(c). The
positive partial transposition (PPT) criterion is a necessary and sufficient
condition for judging the existence of quantum entanglement among $N$\
Gaussian optical beams, when the state has the form of bipartite splitting
with only a single mode on one side like ($1|N-1$) \cite%
{Werner,Adesso,Vollmer}. We characterize the features of quantum
entanglement reduction when an optical mode is transmitted over a lossy
channel with the PPT criterion.

The PPT values are symplectic eigenvalues of a partially transposed matrix.
At the level of quadrature operators, the partial transposition with respect
to mode $k$ ($k$ $=1,2,3$) corresponds to the change of sign of phase
quadrature, $\hat{p}_{k}\longrightarrow -\hat{p}_{k}$. Symplectic
eigenvalues of covariance matrix are defined as positive roots of polynomial 
$\left\vert \gamma ^{T(k)}-i\mu \Omega \right\vert =0$, where $\left\vert
A\right\vert $ denotes the determinant of matrix \cite{Vollmer}. $\gamma
^{T(k)}=T_{k}\gamma T_{k}^{T}$ is the partially transposed matrix of the
quantum state, where $T_{k}$ is a diagonal matrix with all diagonal elements
equal to 1 except for $T_{2k,2k}=-1$, and 
\begin{equation}
\Omega =\oplus _{k=1}^{3}\left( 
\begin{array}{cc}
0 & 1 \\ 
-1 & 0%
\end{array}%
\right) .
\end{equation}%
We consider a bipartite splitting of a three-mode Gaussian state with
covariance matrix $\gamma $ such that one party holds mode $k$ and the other
party possesses the remaining two modes. If the smallest symplectic
eigenvalue $\mu _{k}$ obtained from the polynomial is below 1, the state is
inseparable with respect to the $k|ij$\ splitting.

As shown in Fig. 3, the output optical modes are entangled if the channel
efficiency is larger than 0.24 at the present squeezing level, where the
optimal gain in classical channel is chosen to be $G=0.85$ according to the
requirement of the PPT criterion. If we consider transmission in a fiber
with a loss of 0.2 dB/km, the achievable transmission distance will be about
30 km. Since the optical mode $D_{1}$ comes from the network \emph{B} while $%
D_{2}$ and $D_{3}$ come from the network \emph{A}, the PPT value of ($%
D_{1}|D_{2}D_{3}$) is more sensitive to loss than other two PPT values. The
output entangled state will be more robust against loss in a quantum channel
when the squeezing of resource states increases. For example, when the
squeezing is $-10.9$ dB, which has been realized by H. Vahlbruch\emph{\ et
al.} \cite{Henning}, the obtained entanglement will be quite robust against
loss (dashed-line in Fig. 3). The PPT values in Figs. 3(b) and 3(c) are
smaller than the boundary even when the transmission efficiency in the
quantum channel is zero, and this is because the optical modes $D_{2}$ and $%
D_{3}$ come from the same local network and they are entangled initially.

In summary, we experimentally demonstrate quantum entanglement swapping
between two multipartite GHZ entangled states. After the quantum
entanglement swapping of multipartite entangled states, the quantum modes,
more than two in two multipartite entangled states that never interacted
directly, become entangled. In the experiment, GHZ entangled states are used
as the quantum resources for entanglement swapping. In principle, this
method may also be extended to construct large scale cluster states, which
are very useful for quantum computation. Of course, because quantum
correlation in cluster states is different from that in GHZ states, the
corresponding feedforward scheme needs to be designed according to different
requirements.

The entanglement swapping between a tripartite GHZ state and an EPR
entangled state through a lossy channel equivalent to an optical fiber of 30
km is achieved at present squeezing level. The robustness of the distributed
entanglement over lossy channels depends on the initial squeezing of
multipartite entangled states. Squeezing over 15 dB has been experimentally
generated \cite{Henning}, the use of which will increase the distance of the
entanglement swapping significantly. The robustness of the distributed
entanglement over lossy channels can also be improved by using the existent
techniques. For example, the noiseless linear amplification \cite%
{Ralph2011,Chrzanowski2014,Lvovsky2015}, can be used in the system to
improve the quality of entanglement swapping in a lossy channel. When the
quantum channel is a noisy channel, the noise of which is higher than the
vacuum noise, a correlated noisy channel can be used to remove the effect of
noise on entangled states \cite{Lassen2013}. Since a local quantum network
can be established by distributing a multipartite entangled state to
different quantum nodes, the presented scheme can be used to merge two
space-separated local quantum networks into a large quantum network.

This research was supported by the National Natural Science Foundation of
China (NSFC) (Grants No. 11522433, No. 61475092), the program of Youth
Sanjin Scholar and National Basic Research Program of China (Grant No.
2016YFA0301402).

\section*{APPENDIX}

\subsection{Preparation of the tripartite entangled states}

As shown in Fig. 1 in the maintext, the tripartite
Greenberger-Horne-Zeilinger (GHZ) state A of optical field is prepared by
coupling a phase-squeezed state ($\hat{a}_{2}$) of light and two
amplitude-squeezed states of light ($\hat{a}_{1}$ and $\hat{a}_{3}$) on an
optical beam-splitter network,\ which consists of two optical beam-splitters
with transmittance of $T_{A1}=1/3$ and $T_{A2}=1/2$, respectively. The other
tripartite GHZ state B of optical field is prepared by coupling a
amplitude-squeezed state ($\hat{b}_{3}$) of light and two phase-squeezed
states of light ($\hat{b}_{1}$ and $\hat{b}_{2}$) on an optical
beam-splitter network,\ which consists of two optical beam-splitters with
transmittance of $T_{B1}=1/3$ and $T_{B2}=1/2$, respectively. The
transformation matrixes of the beam-splitter networks for establishing
tripartite GHZ entangled states A and B are given by 
\begin{equation}
U_{A}=\left[ 
\begin{array}{ccc}
\sqrt{\frac{2}{3}} & \sqrt{\frac{1}{3}} & 0 \\ 
-\sqrt{\frac{1}{6}} & \sqrt{\frac{1}{3}} & \sqrt{\frac{1}{2}} \\ 
-\sqrt{\frac{1}{6}} & \sqrt{\frac{1}{3}} & -\sqrt{\frac{1}{2}}%
\end{array}%
\right] ,
\end{equation}%
\begin{equation}
U_{B}=\left[ 
\begin{array}{ccc}
i\sqrt{\frac{2}{3}} & \sqrt{\frac{1}{3}} & 0 \\ 
-i\sqrt{\frac{1}{6}} & \sqrt{\frac{1}{3}} & \sqrt{\frac{1}{2}} \\ 
-i\sqrt{\frac{1}{6}} & \sqrt{\frac{1}{3}} & -\sqrt{\frac{1}{2}}%
\end{array}%
\right] ,
\end{equation}%
respectively. The unitary matrix can be decomposed into a beam-splitter
network $U_{A}=B_{23}^{+}(T_{A2})I_{2}(-1)B_{12}^{+}(T_{A1}),$ $%
U_{B}=B_{23}^{+}(T_{B2})I_{2}(-1)B_{12}^{+}(T_{B1})F_{1},$ where $%
B_{kl}^{+}(T_{j})$ stands for the linearly optical transformation on $j$-th
beam-splitter with transmissivity of $T_{Aj(Bj)}$ ($j=1,2$), $\left(
B_{kl}^{+}\right) _{kk}=\sqrt{1-T},\left( B_{kl}^{+}\right) _{kl}=\left(
B_{kl}^{+}\right) _{lk}=\sqrt{T},\left( B_{kl}^{+}\right) _{ll}=-\sqrt{1-T},$
are matrix elements of the beam-splitter. $I_{k}(-1)=e^{i\pi }$ corresponds
to a $180%
%TCIMACRO{\U{b0}}%
%BeginExpansion
{{}^\circ}%
%EndExpansion
$ rotation in phase space and $F_{1}=e^{i\frac{\pi }{2}}$ corresponds to a
Fourier transformation in phase space. The output modes from the optical
beam-splitter network are expressed by

\begin{eqnarray}
A_{1} &=&\sqrt{\frac{2}{3}}\hat{a}_{1}+\sqrt{\frac{1}{3}}\hat{a}_{2}, \\
A_{2} &=&-\sqrt{\frac{1}{6}}\hat{a}_{1}+\sqrt{\frac{1}{3}}\hat{a}_{2}+\sqrt{%
\frac{1}{2}}\hat{a}_{3,}  \notag \\
A_{3} &=&-\sqrt{\frac{1}{6}}\hat{a}_{1}+\sqrt{\frac{1}{3}}\hat{a}_{2}-\sqrt{%
\frac{1}{2}}\hat{a}_{3},  \notag \\
B_{1} &=&i\sqrt{\frac{2}{3}}\hat{b}_{1}+\sqrt{\frac{1}{3}}\hat{b}_{2}, 
\notag \\
B_{2} &=&-i\sqrt{\frac{1}{6}}\hat{b}_{1}+\sqrt{\frac{1}{3}}\hat{b}_{2}+\sqrt{%
\frac{1}{2}}\hat{b}_{3},  \notag \\
B_{3} &=&-i\sqrt{\frac{1}{6}}\hat{b}_{1}+\sqrt{\frac{1}{3}}\hat{b}_{2}-\sqrt{%
\frac{1}{2}}\hat{b}_{3},  \notag
\end{eqnarray}%
respectively. The quantum correlation noises of two tripartite GHZ states
are given by 
\begin{eqnarray}
\hat{x}_{A_{1}}-\hat{x}_{A_{2}} &=&\sqrt{\frac{3}{2}}\hat{x}%
_{a_{1}}^{(0)}e^{-r}-\sqrt{\frac{1}{2}}\hat{x}_{a_{3}}^{(0)}e^{-r}, \\
\hat{x}_{A_{2}}-\hat{x}_{A_{3}} &=&\sqrt{2}\hat{x}_{a_{3}}^{(0)}e^{-r}, 
\notag \\
\hat{x}_{A_{1}}-\hat{x}_{A_{3}} &=&\sqrt{\frac{3}{2}}\hat{x}%
_{a_{1}}^{(0)}e^{-r}+\sqrt{\frac{1}{2}}\hat{x}_{a_{3}}^{(0)}e^{-r},  \notag
\\
\hat{p}_{A_{1}}+\hat{p}_{A_{2}}+\hat{p}_{A_{3}} &=&\sqrt{3}\hat{p}%
_{a_{2}}^{(0)}e^{-r},  \notag \\
\hat{x}_{B_{1}}-\hat{x}_{B_{2}} &=&-\sqrt{\frac{3}{2}}\hat{p}%
_{b_{1}}^{(0)}e^{-r}-\sqrt{\frac{1}{2}}\hat{x}_{b_{3}}^{(0)}e^{-r},  \notag
\\
\hat{x}_{B_{2}}-\hat{x}_{B_{3}} &=&\sqrt{2}\hat{x}_{b_{3}}^{(0)}e^{-r}, 
\notag \\
\hat{x}_{B_{1}}-\hat{x}_{B_{3}} &=&-\sqrt{\frac{3}{2}}\hat{p}%
_{b_{1}}^{(0)}e^{-r}+\sqrt{\frac{1}{2}}\hat{x}_{b_{3}}^{(0)}e^{-r},  \notag
\\
\hat{p}_{B_{1}}+\hat{p}_{B_{2}}+\hat{p}_{B_{3}} &=&\sqrt{3}\hat{p}%
_{b_{2}}^{(0)}e^{-r},  \notag
\end{eqnarray}%
respectively, where $\hat{x}_{j}^{(0)}$\ and $\hat{p}_{j}^{(0)}$ denote the
quadrature-amplitude and the quadrature-phase operators of corresponding
vacuum field, respectively, and $r$ is the squeezing parameter ($r=0$ and $%
r=+\infty $ correspond to no squeezing and the ideally perfect squeezing,
respectively). Here, we have assumed that six squeezed states have the
identical squeezing parameter. In the experiment, this requirement is easily
achieved by adjusting the three nondegenerate optical parametric amplifiers
(NOPAs) to be operated at the same conditions.

\begin{figure*}[tbp]
\begin{center}
\includegraphics[width=150mm]{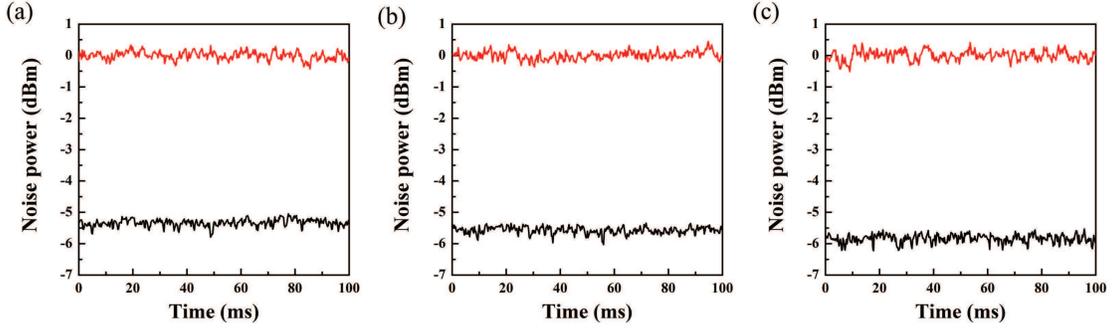}
\end{center}
\caption{\textbf{The measured quantum correlation noises for the prepared
tripartite GHZ entangled state A.} (a) - (c), The measured the correlation
noises are $\Delta ^{2}(x_{A_{1}}-x_{A_{2}})=$ $-5.34\pm 0.11$\ dB, $\Delta
^{2}(\hat{x}_{A_{2}}-\hat{x}_{A_{3}})=-5.57\pm 0.11$ dB, $\Delta ^{2}(\hat{p}%
_{A_{1}}+\hat{p}_{A_{2}}+\hat{p}_{A_{3}})=-5.84\pm 0.13$ dB, respectively.
The red and black lines correspond to the shot noise level and correlation
noises, respectively. Measurement frequency is 3 MHz, parameters of the
spectrum analyzer: resolution bandwidth is 30 kHz, and video bandwidth is
300 Hz.}
\end{figure*}

Fig. 4 shows the experimentally measured quantum correlation noises for the
prepared tripartite GHZ entangled state A. The inseparability criteria for a
tripartite GHZ entangled state is given by \cite{LF}%
\begin{align}
\left\langle \Delta ^{2}(\hat{x}_{A_{1}}-\hat{x}_{A_{2}})\right\rangle
+\left\langle \Delta ^{2}(\hat{p}_{A_{1}}+\hat{p}_{A_{2}}+\hat{p}%
_{A_{3}})\right\rangle & <4,  \label{3} \\
\left\langle \Delta ^{2}(\hat{x}_{A_{2}}-\hat{x}_{A_{3}})\right\rangle
+\left\langle \Delta ^{2}(\hat{p}_{A_{1}}+\hat{p}_{A_{2}}+\hat{p}%
_{A_{3}})\right\rangle & <4.  \notag
\end{align}%
From the measured quantum correlation noises, the calculated values of the
left-hand sides of inequalities (\ref{3}) are $1.37\pm 0.04$ and $1.34\pm
0.04$, respectively, which confirm that the three optical modes are in a
tripartite GHZ entangled state.

\subsection{Entanglement swapping between two tripartite GHZ states}

The optical mode $A_{1}$ is transmitted from multipartite entangled state A
to multipartite entangled state B and mixed with $B_{1}$ on a 1:1
beam-splitter. The output modes $\mu $ and $\nu $ are measured by two
homodyne detectors yield classical photocurrents for the quadratures $\hat{x}%
_{\upsilon }$ and $\hat{p}_{\mu }$, which are%
\begin{eqnarray}
\hat{p}_{\mu } &=&\frac{1}{\sqrt{2}}(\hat{p}_{A_{1}}+\hat{p}_{B_{1}}), \\
\hat{x}_{\upsilon } &=&\frac{1}{\sqrt{2}}(\hat{x}_{A_{1}}-\hat{x}_{B_{1}}), 
\notag
\end{eqnarray}%
respectively. The measurement results of $\sqrt{2}G(\hat{x}_{\upsilon }+\hat{%
p}_{\mu })$ and $\sqrt{2}G\hat{x}_{\upsilon }$ are fed forward to $B_{2}$
and $B_{3}$, respectively. The amplitude and phase quadratures of output
states are expressed by%
\begin{eqnarray}
\hat{x}_{C_{1}} &=&\hat{x}_{A_{3}}, \\
\hat{p}_{C_{1}} &=&\hat{p}_{A_{3}},  \notag \\
\hat{x}_{C_{2}} &=&\hat{x}_{A_{2}},  \notag \\
\hat{p}_{C_{2}} &=&\hat{p}_{A_{2}},  \notag \\
\hat{x}_{C_{3}} &=&\hat{x}_{B_{2}}+\sqrt{2}G\hat{x}_{\upsilon },  \notag \\
\hat{p}_{C_{3}} &=&\hat{p}_{B_{2}}+\sqrt{2}G\hat{p}_{\mu }, \notag \\
\hat{x}_{C_{4}} &=&\hat{x}_{B_{3}}+\sqrt{2}G\hat{x}_{\upsilon }, \notag \\
\hat{p}_{C_{4}} &=&\hat{p}_{B_{3}}, \notag
\end{eqnarray}%
respectively. The parameter $G$ describes the gain in classical channel.

The quantum entanglement among the output modes in multipartite entangled
state C is verified by the inseparability criteria for a four-mode GHZ
entangled state \cite{LF}, which are 
\begin{eqnarray}
\Delta ^{2}(\hat{x}_{C_{1}}-\hat{x}_{C_{2}})+\Delta ^{2}(\hat{p}_{C_{1}}+%
\hat{p}_{C_{2}}+g_{1}\hat{p}_{C_{3}}+g_{2}\hat{p}_{C_{4}}) &<&4,  \notag \\
\Delta ^{2}(\hat{x}_{C_{2}}-\hat{x}_{C_{3}})+\Delta ^{2}(g_{3}\hat{p}%
_{C_{1}}+\hat{p}_{C_{2}}+\hat{p}_{C_{3}}+g_{4}\hat{p}_{C_{4}}) &<&4,  \notag
\\
\Delta ^{2}(\hat{x}_{C_{3}}-\hat{x}_{C_{4}})+\Delta ^{2}(g_{5}\hat{p}%
_{C_{1}}+g_{6}\hat{p}_{C_{2}}+\hat{p}_{C_{3}}+\hat{p}_{C_{4}}) &<&4,  \notag
\\
&&  \label{4}
\end{eqnarray}%
where $g_{i}$ $(i=1,2,...,6)$ is the optimal gain used to minimize the
correlation variances at the left-hand sides of Eq. (\ref{4}).

The correlation variances of quadrature components among\ the output states
are expressed by%
\begin{eqnarray}
V_{1} &=&\left\langle \Delta ^{2}(\hat{x}_{C_{1}}-\hat{x}_{C_{2}})\right%
\rangle =2V, \\
V_{2} &=&\left\langle \Delta ^{2}(\hat{p}_{C_{1}}+\hat{p}_{C_{2}}+g_{1}\hat{p%
}_{C_{3}}+g_{2}\hat{p}_{C_{4}})\right\rangle  \notag \\
&=&\frac{[(2+Gg_{1})^{2}+(g_{1}+Gg_{1}+g_{2})^{2}]V}{3}  \notag \\
&&+\frac{[4(Gg_{1}-1)^{2}+3(g_{1}-g_{2})^{2}+(g_{1}-2Gg_{1}+g_{2})^{2}]V^{%
\prime }}{6},  \notag \\
V_{3} &=&\left\langle \Delta ^{2}(\hat{x}_{C_{2}}-\hat{x}_{C_{3}})\right%
\rangle  \notag \\
&=&\frac{2[(G-1)^{2}V^{\prime }+2(1+G+G^{2})V]}{3},  \notag \\
V_{4} &=&\left\langle \Delta ^{2}(g_{3}\hat{p}_{C_{1}}+\hat{p}_{C_{2}}+\hat{p%
}_{C_{3}}+g_{4}\hat{p}_{C_{4}})\right\rangle  \notag \\
&=&\frac{[(1+G+g_{3})^{2}+(1+g_{4}+G)^{2}]V}{3}  \notag \\
&&+\frac{[(2G-1-g_{3})^{2}+3(1-g_{3})^{2}]V^{\prime }}{6}  \notag \\
&&+\frac{(2G-g_{4}-1)^{2}+3(1-g_{4})^{2}]V^{\prime }}{6},  \notag \\
V_{5} &=&\left\langle \Delta ^{2}(\hat{x}_{C_{3}}-\hat{x}_{C_{4}})\right%
\rangle =2V,  \notag \\
V_{6} &=&\left\langle \Delta ^{2}(g_{5}\hat{p}_{C_{1}}+g_{6}\hat{p}_{C_{2}}+%
\hat{p}_{C_{3}}+\hat{p}_{C_{4}})\right\rangle  \notag \\
&=&\frac{[(G+g_{5}+g_{6})^{2}+(2+G)^{2}]V}{3}  \notag \\
&&+\frac{[(2G-g_{5}-g_{6})^{2}+3(g_{6}-g_{5})^{2}+(2G-2)^{2}]V^{\prime }}{6},
\notag
\end{eqnarray}%
respectively, where $V=e^{-2r}$ and $V^{\prime }=e^{2r}$ represent the
variances of squeezed and anti-squeezed quadratures of the optical mode,
respectively. The optimal gain in classical channel equals to 
\begin{equation}
G=\frac{V^{\prime }-V}{V^{\prime }+V}.
\end{equation}%
The calculated optimal gain $g_{i}$ $(i=1,2,...,6)$ are 
\begin{eqnarray}
g_{1} &=&\frac{2(V^{\prime }-V)^{2}(V^{\prime }+V)(2V^{\prime }+V)}{%
4V^{\prime 4}+14V^{\prime 3}V+9V^{\prime 2}V^{2}+8V^{\prime }V^{3}+V^{4}},
\label{g1} \\
g_{2} &=&\frac{4V^{\prime }(V^{\prime }-V)^{3}}{4V^{\prime 4}+14V^{\prime
3}V+9V^{\prime 2}V^{2}+8V^{\prime }V^{3}+V^{4}},  \notag \\
\end{eqnarray}
\begin{eqnarray}
g_{3} &=&\frac{2(V^{\prime 2}-V^{\prime }V)}{(V^{\prime }+V)(2V^{\prime }+V)}%
,  \notag \\
g_{4} &=&\frac{2(V^{\prime 2}-V^{\prime }V)}{(V^{\prime }+V)(2V^{\prime }+V)}%
,  \notag \\
g_{5} &=&\frac{(V^{\prime }-V)^{2}}{(V^{\prime }+V)(V^{\prime }+2V)},  \notag
\\
g_{6} &=&\frac{(V^{\prime }-V)^{2}}{(V^{\prime }+V)(V^{\prime }+2V)}.  \notag
\end{eqnarray}

\begin{figure*}[tbp]
\begin{center}
\includegraphics[width=150mm]{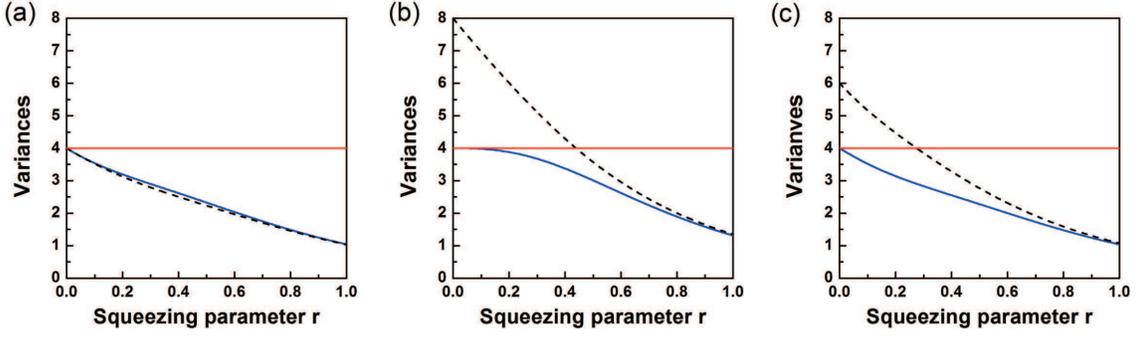}
\end{center}
\caption{\textbf{The inseparability criteria for the obtained four-mode GHZ
entangled state.} (a) - (c) are corresponding to the first, second and third
inequalities in Eq. (\protect\ref{4}), respectively. The dashed and solid
lines correspond to the case of unit and optimal gain factor in the
classical channel, respectively.}
\end{figure*}

Fig. 5 shows the dependence of inseparability criteria of four-mode GHZ
entangled state on squeezing parameter. The optimal gains $g_{i}$ $%
(i=1,2,...,6)$ in Eq. (\ref{g1}) are chosen in the calculation. If the unit
gain is chosen, the obtained state is entangled only when the squeezing
parameter is higher than $0.44$ ($-3.82$ dB squeezing, dashed lines).
However, when optimal gain factor $G=0.95$ is used, the requirement of
entanglement on the squeezing parameter is reduced (solid lines).

\subsection{Entanglement swapping between a tripartite GHZ state and an EPR
state}

The EPR entangled state is prepared by coupling a phase-squeezed state ($%
\hat{b}_{1}$) of light and a amplitude-squeezed state of light ($\hat{b}_{2}$%
) on a 1:1 beam-splitter, the output states are 
\begin{eqnarray}
E_{1} &=&\frac{1}{\sqrt{2}}(\hat{b}_{1}+\hat{b}_{2}), \\
E_{2} &=&\frac{1}{\sqrt{2}}(\hat{b}_{1}-\hat{b}_{2}),  \notag
\end{eqnarray}%
respectively. The quantum correlations between the amplitude and phase
quadratures of the EPR entangled state are $\Delta ^{2}\left( \hat{x}%
_{E_{1}}-\hat{x}_{E_{2}}\right) =$ $\Delta ^{2}\left( \hat{p}_{E_{1}}+\hat{p}%
_{E_{2}}\right) =2e^{-2r}$.

\begin{figure*}[tbp]
\begin{center}
\includegraphics[width=120mm]{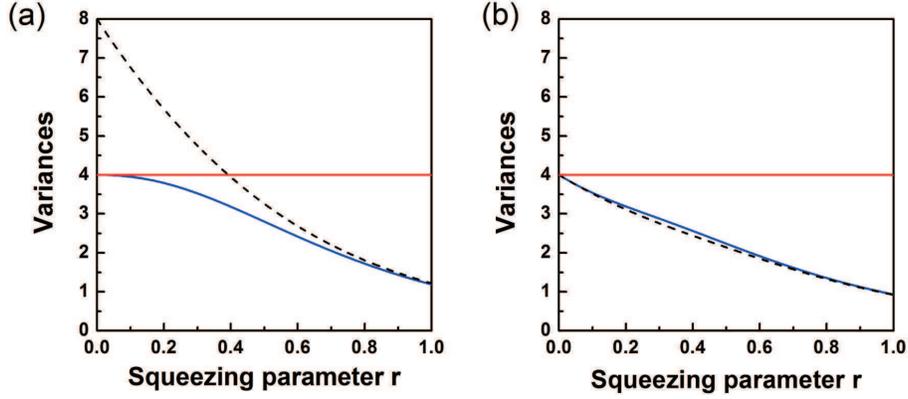}
\end{center}
\caption{\textbf{The inseparability criteria for the obtained tripartite GHZ
entangled state.} (a) and (b) correspond to the first and second
inequalities in Eq. (\protect\ref{3b}), respectively. The dashed and solid
lines correspond to the case of unit and optimal gain factor in the
classical channel, respectively.}
\end{figure*}

After the transmission of optical mode $\hat{A}_{1}$\ over a lossy channel,
the output mode is given by $\hat{A}_{L}=\sqrt{\eta }\hat{A}_{1}+\sqrt{%
1-\eta }\hat{\nu}_{0}$, where $\eta $\ and $\hat{\nu}_{0}$\ represent the
transmission efficiency of quantum channel and vacuum state induced by loss
into the quantum channel, respectively. Then modes $A_{L}$ and $E_{1}$ are
combined by a 1:1 beam splitter, the output modes of the beam-splitter are
measured by two homodyne detectors yield classical photocurrents for the
quadratures $\hat{x}_{\upsilon }$ and $\hat{p}_{\mu }$, which are 
\begin{eqnarray}
\hat{p}_{\mu } &=&\frac{1}{\sqrt{2}}(\hat{p}_{A_{L}}+\hat{p}_{E_{1}}), \\
\hat{x}_{\upsilon } &=&\frac{1}{\sqrt{2}}(\hat{x}_{A_{L}}-\hat{x}_{E_{1}}), 
\notag
\end{eqnarray}%
respectively. The measurement results of $\sqrt{2}G\hat{x}_{\upsilon }$ and $%
\sqrt{2}G\hat{p}_{\mu }$ are fed forward to $E_{2}$. The quadratures of
output states are expressed by%
\begin{eqnarray}
\hat{x}_{D_{1}} &=&\hat{x}_{E_{2}}+\sqrt{2}G\hat{x}_{\upsilon }, \\
\hat{p}_{D_{1}} &=&\hat{p}_{E_{2}}+\sqrt{2}G\hat{p}_{\mu },  \notag \\
\hat{x}_{D_{2}} &=&\hat{x}_{A_{2}},  \notag \\
\hat{p}_{D_{2}} &=&\hat{p}_{A_{2}},  \notag \\
\hat{x}_{D_{3}} &=&\hat{x}_{A_{3}},  \notag \\
\hat{p}_{D_{3}} &=&\hat{p}_{A_{3}},  \notag
\end{eqnarray}%
respectively.

The inseparability criteria for a three-mode GHZ entangled state established
in the entanglement swapping between a tripartite GHZ state and an EPR
entangled state are given by \cite{LF} 
\begin{eqnarray}
\Delta ^{2}(\hat{x}_{D_{1}}-\hat{x}_{D_{2}})+\Delta ^{2}(\hat{p}_{D_{1}}+%
\hat{p}_{D_{2}}+g_{7}\hat{p}_{D_{3}}) &<&4,  \label{3b} \\
\Delta ^{2}(\hat{x}_{D_{2}}-\hat{x}_{D_{3}})+\Delta ^{2}(g_{8}\hat{p}%
_{D_{1}}+\hat{p}_{D_{2}}+\hat{p}_{D_{3}}) &<&4,  \notag
\end{eqnarray}%
where $g_{j}$ $(j=7,8)$ is the optimal gain used to minimize the correlation
variances at the left-hand sides of Eq. (\ref{3b}). When $\eta =1,$ the
correlation variances of quadrature components among the output states are
expressed by%
\begin{eqnarray}
V_{7} &=&\left\langle \Delta ^{2}(\hat{x}_{D_{1}}-\hat{x}_{D_{2}})\right%
\rangle \\
&=&\frac{[(1+2G)^{2}+3+3(1+G)^{2}]V}{6}  \notag \\
&&+\frac{[2(-1+G)^{2}+3(1-G)^{2}]V^{\prime }}{6},  \notag \\
V_{8} &=&\left\langle \Delta ^{2}(\hat{p}_{D_{1}}+\hat{p}_{D_{2}}+g_{7}\hat{p%
}_{D_{3}})\right\rangle  \notag \\
&=&\frac{[2(1+G+g_{7})^{2}+3(G+1)^{2}]V}{6}  \notag \\
&&+\frac{[(2G-1-g_{7})^{2}+3(1-g_{7})^{2}+3(G-1)^{2}]V^{\prime }}{6},  \notag \\
V_{9} &=&\left\langle \Delta ^{2}(\hat{x}_{D_{2}}-\hat{x}_{D_{3}})\right%
\rangle =2V,  \notag \\
V_{10} &=&\left\langle \Delta ^{2}(g_{8}\hat{p}_{D_{1}}+\hat{p}_{D_{2}}+\hat{%
p}_{D_{3}})\right\rangle  \notag \\
&=&\frac{[2(2+Gg_{8})^{2}+3(Gg_{8}+g_{8})^{2}]V}{6}  \notag \\
&&+\frac{[4(Gg_{8}-1)^{2}+3(Gg_{8}-g_{8})^{2}]V^{\prime }}{6}.  \notag
\end{eqnarray}%
\newline
When $\eta =1,$ the optimal gain in classical channel is expressed by 
\begin{equation}
G=\frac{V^{\prime }-V}{V^{\prime }+V}.
\end{equation}%
The optimal gain $g_{j}$ $(j=7,8)$ to minimize the left-hand sides of the
inequalities are 
\begin{eqnarray}
g_{7} &=&\frac{2V^{\prime }(V^{\prime }-V)}{(V^{\prime }+V)(2V^{\prime }+V)},
\label{g7} \\
g_{8} &=&\frac{2(V^{\prime }-V)^{2}(V^{\prime }+V)}{2V^{\prime 3}+3V^{\prime
2}V+6V^{\prime }V^{2}+V^{3}}.  \notag
\end{eqnarray}

Fig. 6 shows the dependence of inseparability criteria of tripartite GHZ
entangled state on squeezing parameter. The optimal gains $g_{j}$ $(j=7,8)$
in Eq. (\ref{g7}) are chosen in the calculation. If the unit gain is chosen,
the obtained state is entangled only when the squeezing parameter is higher
than $0.39$ ($-3.41$ dB squeezing, dashed lines). When the optimal gain
factor\ $G=0.95$ is chosen, the requirement to the squeezing level is
reduced for achieving resultant entangled state (solid lines).

\subsection{Covariance matrix of the output state}

Gaussian state is the state with Gaussian characteristic functions and
quasi-probability distributions on the multi-mode quantum phase space, which
can be completely characterized by a covariance matrix. The elements of the
covariance matrix are $\sigma _{ij}=Cov\left( \hat{R}_{i},\hat{R}_{j}\right)
=\frac{1}{2}\left\langle \hat{R}_{i}\hat{R}_{j}+\hat{R}_{j}\hat{R}%
_{i}\right\rangle -\left\langle \hat{R}_{i}\right\rangle \left\langle \hat{R}%
_{j}\right\rangle $, where $\hat{R}=(\hat{x}_{D_{1}},\hat{p}_{D_{1}},\hat{x}%
_{D_{2}},\hat{p}_{D_{2}},\hat{x}_{D_{3}},\hat{p}_{D_{3}})^{T}$ is a vector
composed by the amplitude and phase quadratures of tripartite optical beams.
Thus the covariance matrix of the tripartite optical beams is expressed as%
\begin{equation}
\sigma =\left[ 
\begin{array}{ccc}
\sigma _{D_{1}\text{ }} & \sigma _{_{D_{1}D_{2}}} & \sigma _{_{D_{1}D_{3}}}
\\ 
\sigma _{D_{1}D_{2}}^{T} & \sigma _{D_{2}\text{ }} & \sigma _{_{D_{2}D_{3}}}
\\ 
\sigma _{D_{1}D_{3}}^{T} & \sigma _{D_{2}D_{3}}^{T} & \sigma _{D_{3}\text{ }}%
\end{array}%
\right] ,
\end{equation}%
where%
\begin{eqnarray}
\sigma _{D_{1}\text{ }} &=&\left[ 
\begin{array}{cc}
\bigtriangleup ^{2}\hat{x}_{D_{1}} & 0 \\ 
0 & \bigtriangleup ^{2}\hat{p}_{D_{1}}%
\end{array}%
\right] , \\
\sigma _{D_{2}\text{ }} &=&\left[ 
\begin{array}{cc}
\bigtriangleup ^{2}\hat{x}_{D_{2}} & 0 \\ 
0 & \bigtriangleup ^{2}\hat{p}_{D_{2}}%
\end{array}%
\right] ,  \notag \\
\sigma _{D_{3}\text{ }} &=&\left[ 
\begin{array}{cc}
\bigtriangleup ^{2}\hat{x}_{D_{3}} & 0 \\ 
0 & \bigtriangleup ^{2}\hat{p}_{D_{3}}%
\end{array}%
\right] ,  \notag \\
\sigma _{_{D_{1}D_{2}}} &=&\left[ 
\begin{array}{cc}
Cov\left( \hat{x}_{D_{1}},\hat{x}_{D_{2}}\right) & Cov\left( \hat{x}_{D_{1}},%
\hat{p}_{D_{2}}\right) \\ 
Cov\left( \hat{p}_{D_{1}},\hat{x}_{D_{2}}\right) & Cov\left( \hat{p}_{D_{1}},%
\hat{p}_{D_{2}}\right)%
\end{array}%
\right] ,  \notag \\
\sigma _{_{D_{1}D_{3}}} &=&\left[ 
\begin{array}{cc}
Cov\left( \hat{x}_{D_{1}},\hat{x}_{D_{3}}\right) & Cov\left( \hat{x}_{D_{1}},%
\hat{p}_{D_{3}}\right) \\ 
Cov\left( \hat{p}_{D_{1}},\hat{x}_{D_{3}}\right) & Cov\left( \hat{p}_{D_{1}},%
\hat{p}_{D_{3}}\right)%
\end{array}%
\right] ,  \notag \\
\sigma _{_{D_{2}D_{3}}} &=&\left[ 
\begin{array}{cc}
Cov\left( \hat{x}_{D_{2}},\hat{x}_{D_{3}}\right) & Cov\left( \hat{x}_{D_{2}},%
\hat{p}_{D_{3}}\right) \\ 
Cov\left( \hat{p}_{D_{2}},\hat{x}_{D_{3}}\right) & Cov\left( \hat{p}_{D_{2}},%
\hat{p}_{D_{3}}\right)%
\end{array}%
\right] ,  \notag
\end{eqnarray}%
respectively.

For the output modes $D_{1}$, $D_{2}$ and $D_{3}$ in a lossy channel, we have

\begin{eqnarray}
\bigtriangleup ^{2}\hat{x}_{D_{1}} &=&\frac{4G^{2}\eta +3(1+G)^{2}}{6}V \\
&&+\frac{2G^{2}\eta +3(1-G)^{2}}{6}V^{\prime }+G^{2}(1-\eta ),  \notag \\
\bigtriangleup ^{2}\hat{p}_{D_{1}} &=&\frac{2G^{2}\eta +3(1+G)^{2}}{6}V 
\notag \\
&&+\frac{4G^{2}\eta +3(G-1)^{2}}{6}V^{\prime }+G^{2}(1-\eta ),  \notag \\
\bigtriangleup ^{2}\hat{x}_{D_{2}} &=&\bigtriangleup ^{2}\hat{x}_{D_{3}}=%
\frac{2}{3}V+\frac{1}{3}V^{\prime },  \notag \\
\bigtriangleup ^{2}\hat{p}_{D_{2}} &=&\bigtriangleup ^{2}\hat{p}_{D_{3}}=%
\frac{1}{3}V+\frac{2}{3}V^{\prime },  \notag \\
Cov\left( \hat{x}_{D_{1}},\hat{x}_{D_{2}}\right) &=&Cov\left( \hat{x}%
_{D_{1}},\hat{x}_{D_{3}}\right) =-\frac{G\sqrt{\eta }}{3}V+\frac{G\sqrt{\eta 
}}{3}V^{\prime },  \notag \\
Cov\left( \hat{p}_{D_{1}},\hat{p}_{D_{2}}\right) &=&Cov\left( \hat{p}%
_{D_{1}},\hat{p}_{D_{3}}\right) =-\frac{G\sqrt{\eta }}{3}V^{\prime }+\frac{G%
\sqrt{\eta }}{3}V,  \notag 
\end{eqnarray}
\begin{eqnarray}
Cov\left( \hat{x}_{D_{2}},\hat{x}_{D_{3}}\right) &=&-\frac{1}{3}V+\frac{1}{3}%
V^{\prime },  \notag \\
Cov\left( \hat{p}_{D_{2}},\hat{p}_{D_{3}}\right) &=&-\frac{1}{3}V^{\prime }+%
\frac{1}{3}V,  \notag \\
Cov\left( \hat{x}_{D_{1}},\hat{p}_{D_{2}}\right) &=&Cov\left( \hat{p}%
_{D_{1}},\hat{x}_{D_{2}}\right) =0,  \notag \\
Cov\left( \hat{x}_{D_{1}},\hat{p}_{D_{3}}\right) &=&Cov\left( \hat{p}%
_{D_{1}},\hat{x}_{D_{3}}\right) =0,  \notag \\
Cov\left( \hat{x}_{D_{2}},\hat{p}_{D_{3}}\right) &=&Cov\left( \hat{p}%
_{D_{2}},\hat{x}_{D_{3}}\right) =0.  \notag
\end{eqnarray}

To partially reconstruct all relevant entries of its associated covariance
matrix, we have performed 18 different measurements on the output optical
modes. These measurements include the variances of the amplitude and phase
quadratures of three output optical modes, and the cross correlation
variances $\Delta ^{2}\left( \hat{x}_{D_{1}}-\hat{x}_{D_{2}}\right) $, $%
\Delta ^{2}\left( \hat{x}_{D_{1}}-\hat{x}_{D_{3}}\right) $, $\Delta
^{2}\left( \hat{x}_{D_{2}}-\hat{x}_{D_{3}}\right) $, $\Delta ^{2}\left( \hat{%
p}_{D_{1}}+\hat{p}_{D_{2}}\right) $, $\Delta ^{2}\left( \hat{p}_{D_{1}}+\hat{%
p}_{D_{3}}\right) $, $\Delta ^{2}\left( \hat{p}_{D_{2}}+\hat{p}%
_{D_{3}}\right) $, $\Delta ^{2}\left( \hat{x}_{D_{1}}+\hat{p}_{D_{2}}\right) 
$, $\Delta ^{2}\left( \hat{x}_{D_{2}}+\hat{p}_{D_{1}}\right) $, $\Delta
^{2}\left( \hat{x}_{D_{1}}+\hat{p}_{D_{3}}\right) $, $\Delta ^{2}\left( \hat{%
x}_{D_{3}}+\hat{p}_{D_{1}}\right) $, $\Delta ^{2}\left( \hat{x}_{D_{2}}+\hat{%
p}_{D_{3}}\right) $ and $\Delta ^{2}\left( \hat{x}_{D_{3}}+\hat{p}%
_{D_{2}}\right) $, respectively. The covariance elements are calculated via 
\cite{Steinlechner} 
\begin{align}
Cov\left( \hat{R}_{i},\hat{R}_{j}\right) & =\frac{1}{2}\left[ \Delta
^{2}\left( \hat{R}_{i}+\hat{R}_{j}\right) -\Delta ^{2}\hat{R}_{i}-\Delta ^{2}%
\hat{R}_{j}\right] , \\
Cov\left( \hat{R}_{i},\hat{R}_{j}\right) & =-\frac{1}{2}\left[ \Delta
^{2}\left( \hat{R}_{i}-\hat{R}_{j}\right) -\Delta ^{2}\hat{R}_{i}-\Delta ^{2}%
\hat{R}_{j}\right] .  \notag
\end{align}%
In the experiment, we obtain all the covariance matrices of every quantum
state actually, and then calculate the PPT eigenvalues to verify whether the
quantum states are entangled or not.

In our experiment, the squeezing and anti-squeezing noises of squeezed
states are $-5.90$ dB and $9.84$ dB, which correspond to $V=0.26$ and $%
V^{\prime }=9.64$, respectively. The optimal gain $G=0.85$ is chosen to
optimize PPT eigenvalues. We measured covariance matrix of the output modes $%
D_{1}$, $D_{2}$ and $D_{3}$ with different transmission efficiency $\eta
=0.98,0.80,0.60,0.40$ and $0.20$, the reconstructed covariance matrix are 
\begin{equation}
\sigma _{1}=\left( 
\begin{array}{cccccc}
3.05 & 0 & 2.74 & 0.21 & 2.67 & 0.07 \\ 
0 & 5.34 & -0.20 & -2.54 & -0.15 & -2.88 \\ 
2.74 & -0.20 & 3.46 & 0 & 3.13 & 0.01 \\ 
0.21 & -2.54 & 0 & 6.37 & 0.25 & -2.94 \\ 
2.67 & -0.15 & 3.13 & 0.25 & 3.36 & 0 \\ 
0.07 & -2.88 & 0.01 & -2.94 & 0 & 6.68%
\end{array}%
\right) ,
\end{equation}%
\begin{equation}
\sigma _{2}=\left( 
\begin{array}{cccccc}
2.68 & 0 & 2.44 & 0.11 & 2.50 & 0.11 \\ 
0 & 4.75 & -0.08 & -2.42 & -0.18 & -2.51 \\ 
2.44 & -0.08 & 3.51 & 0 & 3.31 & 0.13 \\ 
0.11 & -2.42 & 0 & 6.44 & 0.02 & -3.08 \\ 
2.50 & -0.18 & 3.31 & 0.02 & 3.67 & 0 \\ 
0.11 & -2.51 & 0.13 & -3.08 & 0 & 6.53%
\end{array}%
\right) ,
\end{equation}%
\begin{equation}
\sigma _{3}=\left( 
\begin{array}{cccccc}
2.39 & 0 & 2.17 & 0.15 & 2.24 & -0.03 \\ 
0 & 3.89 & 0.008 & -1.97 & -0.13 & -2.24 \\ 
2.17 & 0.008 & 3.51 & 0 & 3.31 & 0.06 \\ 
0.15 & -1.97 & 0 & 6.44 & 0.02 & -3.16 \\ 
2.24 & -0.13 & 3.31 & 0.02 & 3.67 & 0 \\ 
-0.03 & -2.24 & 0.06 & -3.16 & 0 & 6.68%
\end{array}%
\right) ,
\end{equation}%
\begin{equation}
\sigma _{4}=\left( 
\begin{array}{cccccc}
1.92 & 0 & 1.72 & 0.42 & 1.80 & 0.30 \\ 
0 & 2.93 & -0.03 & -1.57 & -0.07 & -1.60 \\ 
1.72 & -0.03 & 3.51 & 0 & 3.31 & 0.06 \\ 
0.42 & -1.57 & 0 & 6.44 & 0.02 & -3.16 \\ 
1.80 & -0.07 & 3.31 & 0.02 & 3.67 & 0 \\ 
0.30 & -1.60 & 0.06 & -3.16 & 0 & 6.68%
\end{array}%
\right) ,
\end{equation}%
\begin{equation}
\sigma _{5}=\left( 
\begin{array}{cccccc}
1.56 & 0 & 1.20 & 0.27 & 1.29 & 0.28 \\ 
0 & 2.23 & -0.16 & -1.16 & -0.18 & -1.17 \\ 
1.20 & -0.16 & 3.51 & 0 & 3.31 & 0.13 \\ 
0.27 & -1.16 & 0 & 6.44 & 0.02 & -3.08 \\ 
1.29 & -0.18 & 3.31 & 0.02 & 3.67 & 0 \\ 
0.28 & -1.17 & 0.13 & -3.08 & 0 & 6.53%
\end{array}%
\right) ,
\end{equation}%
respectively. The entanglement among the prepared tripartite state is
evaluated by PPT criterion and we obtain PPT eigenvalues PPT$_{\text{D}_{%
\text{1}}}=0.52$, PPT$_{\text{D}_{\text{2}}}=0.39$, PPT$_{\text{D}_{\text{3}%
}}=0.40$ for $\eta =0.98$, and PPT$_{\text{D}_{\text{1}}}=0.61$, PPT$_{\text{%
D}_{\text{2}}}=0.42$, PPT$_{\text{D}_{\text{3}}}=0.42$ for $\eta =0.80$, and
PPT$_{\text{D}_{\text{1}}}=0.74$, PPT$_{\text{D}_{\text{2}}}=0.50$, PPT$_{%
\text{D}_{\text{3}}}=0.50$ for $\eta =0.60$, and PPT$_{\text{D}_{\text{1}%
}}=0.86$, PPT$_{\text{D}_{\text{2}}}=0.58$, PPT$_{\text{D}_{\text{3}}}=0.56$
for $\eta =0.40$, and PPT$_{\text{D}_{\text{1}}}=1.03$, PPT$_{\text{D}_{%
\text{2}}}=0.70$, PPT$_{\text{D}_{\text{3}}}=0.66$ for $\eta =0.20$,
respectively.


\begin{thebibliography}{99}
\bibitem{Briegel} H. J. Briegel and R. Raussendorf, Phys. Rev. Lett. \textbf{%
86,} 910 (2001).

\bibitem{Zhang2006} J. Zhang and S. L. Braunstein, Phys. Rev. A \textbf{73,}
032318 (2006).

\bibitem{Raussendorf2001} R. Raussendorf and H. J. Briegel, Phys. Rev. Lett. 
\textbf{86,} 5188 (2001).

\bibitem{Menicucci2006} N. C. Menicucci\textit{, }P. van Loock\textit{, }M.
Gu, C. Weedbrook, T. C. Ralph, and M. A. Nielsen, Phys. Rev. Lett. \textbf{%
97,} 110501 (2006).

\bibitem{Ukai2011} R. Ukai, N. Iwata, Y. Shimokawa, S. C. Armstrong, A.
Politi, J. I. Yoshikawa, P. van Loock, and A. Furusawa, Phys. Rev. Lett. 
\textbf{106,} 240504 (2011).

\bibitem{Su2013} X. Su, S. Hao, X. Deng, L. Ma, M. Wang, X. Jia, C. Xie, and
K. Peng,\textit{\ }Nat. Commun. \textbf{4,} 2828 (2013).

\bibitem{Loock} P. van Loock and S. L. Braunstein, Phys. Rev. Lett. \textbf{%
84,} 3482 (2000).

\bibitem{Hide} H. Yonezawa, T. Aoki, and A. Furusawa, Nature (London) 
\textbf{431,} 430 (2004).

\bibitem{Jing} J. Jing, J. Zhang, Y. Yan, F. Zhao, C. Xie, and K. Peng,
Phys. Rev. Lett. \textbf{90,} 167903 (2003).

\bibitem{Jona} J. Roslund, R. M. de Ara\'{u}jo, S. Jiang, C. Fabre, and N.
Treps, Nat. Photonics \textbf{8,} 109 (2014).

\bibitem{Browne2005} D. E. Browne and T. Rudolph, Phys. Rev. Lett. \textbf{%
95,} 010501 (2005).

\bibitem{Miwa} Y. Miwa, R. Ukai, J.-I. Yoshikawa, R. Filip, P. van Loock,
and A. Furusawa, Phys. Rev. A \textbf{82,} 032305 (2010).

\bibitem{Bose} S. Bose, V. Vedral, and P. L. Knight, Phys. Rev. A \textbf{57,%
} 822 (1998).

\bibitem{Komar} P. K\'{o}m\'{a}r, E. M. Kessler, M. Bishof, L. Jiang, A. S.
S\/{\o r}ensen, J. Ye, and M. D. Lukin, Nat. Phys. \textbf{10,} 582 (2014).

\bibitem{Pan1} J. Yin \textit{et al.} Nature (London) \textbf{488,} 185
(2012).

\bibitem{Ma} X.-S. Ma \textit{et al.} Nature (London) \textbf{489,} 269
(2012).

\bibitem{Nam} H. Takesue, S. D. Dyer, M. J. Stevens, V. Verma, R. P. Mirin,
and S.W. Nam, Optica \textbf{2,} 832 (2015).

\bibitem{Zukowski} M. \.{Z}ukowski, A. Zeilinger, M. A. Horne, and A. K.
Ekert, Phys. Rev. Lett. \textbf{71,} 4287 (1993).

\bibitem{Pan} J.-W. Pan, D. Bouwmeester, H. Weinfurter, and A. Zeilinger,
Phys. Rev. Lett. \textbf{80,} 3891 (1998).

\bibitem{Sciarrino} F. Sciarrino, E. Lombardi, G. Milani, and F. De Martini,
Phys. Rev. A \textbf{66,} 024309 (2002).

\bibitem{Riedmatten} H. de Riedmatten, I. Marcikic, J. A. W. van
Houwelingen, W. Tittel, H. Zbinden, and N. Gisin, Phys. Rev. A \textbf{71,}
050302(R) (2005).

\bibitem{Ralph} R. E. S. Polkinghorne and T. C. Ralph, Phys. Rev. Lett. 83,
2095 (1999).

\bibitem{Tan} S. M. Tan, Phys. Rev. A \textbf{60,} 2752 (1999).

\bibitem{Loock2} P. van Loock and S. L. Braunstein, Phys. Rev. A \textbf{61,}
010302(R) (1999).

\bibitem{Jia} X. Jia, X. Su, Q. Pan, J. Gao, C. Xie, and K. Peng, Phys. Rev.
Lett. \textbf{93,} 250503 (2004).

\bibitem{Takei} N. Takei, H. Yonezawa, T. Aoki, and A. Furusawa, Phys. Rev.
Lett. \textbf{94,} 220502 (2005).

\bibitem{Takeda} S. Takeda, M. Fuwa, P. van Loock, and A. Furusawa, Phys.
Rev. Lett. \textbf{114,} 100501 (2015).

\bibitem{Andersen} U. L. Andersen, J. S. Neergaard-Nielsen, P. van Loock,
and A. Furusawa, Nat. Phys. \textbf{11, }713 (2015).

\bibitem{Lu2009} C.-Y. Lu, T. Yang, and J.-W. Pan, Phys. Rev. Lett. \textbf{%
103,} 020501 (2009).

\bibitem{Zhou} Y. Zhou, X. Jia, F. Li, C. Xie, and K. Peng, Opt. Express, 
\textbf{23,} 4952 (2015).

\bibitem{Su2007} X. Su, A. Tan, X. Jia, J. Zhang, C. Xie, and K. Peng, Phys.
Rev. Lett. \textbf{98,} 070502 (2007).

\bibitem{LF} P. van Loock and A. Furusawa, Phys. Rev. A \textbf{67,} 052315
(2003).

\bibitem{Werner} R. F. Werner and M. M. Wolf, Phys. Rev. Lett. \textbf{86,}
3658 (2001).

\bibitem{Adesso} G. Adesso, A. Serafini, and F. Illuminati, Phys. Rev. A%
\textit{\ }\textbf{73, }032345 (2006).

\bibitem{Vollmer} C. E. Vollmer, D. Schulze, T. Eberle, V. H\"{a}ndchen, J.
Fiur\'{a}\v{s}ek, and R\textit{. }Schnabel, Phys. Rev. Lett. \textbf{111, }%
230505 (2013).

\bibitem{Henning} H. Vahlbruch, M. Mehmet, K. Danzmann, and R. Schnabel%
\textit{. }Phys. Rev. Lett. \textbf{117, }110801 (2016).

\bibitem{Ralph2011} T. C. Ralph, Phys. Rev. A \textbf{84,} 022339 (2011).

\bibitem{Chrzanowski2014} H. M. Chrzanowski, N. Walk, S. M. Assad, J.
Janousek, S. Hosseini, T. C. Ralph, T. Symul, and P. K. Lam, Nat. Photonics 
\textbf{8,} 333 (2014).

\bibitem{Lvovsky2015} A. E. Ulanov, I. A. Fedorov, A. A. Pushkina, Y. V.
Kurochkin, T. C. Ralph, and A. I. Lvovsky, Nat. Photonics \textbf{9,} 764
(2015).

\bibitem{Lassen2013} M. Lassen, A. Berni, L. S. Madsen, R. Filip, and U. L.
Andersen, Phys. Rev. Lett. \textbf{111,} 180502 (2013).

\bibitem{Steinlechner} S. Steinlechner, J. Bauchrowitz, T. Eberle, and R.
Schnabel, Phys. Rev. A \textbf{87,} 022104 (2013).
\end{thebibliography}
\end{document}